# VOLUMETRIC AND VISCOSITY DATA OF 1-IODONAPHTHALENE + *n*-ALKANE MIXTURES AT (288.15-308.15) K

Luis Felipe Sanz, Juan Antonio González[*], Fernando Hevia, Daniel Lozano-Martín, João Victor Alves-Laurentino, Fatemeh Pazoki, Isaías García de la Fuente, José Carlos Cobos

G.E.T.E.F., Departamento de Física Aplicada, Facultad de Ciencias, Universidad de Valladolid, Paseo de Belén, 7, 47011 Valladolid, Spain

*Corresponding author, e-mail: jagl@termo.uva.es

**Abstract**

Density and viscosity measurements have been performed for the systems 1-iodonaphthalene + heptane, or + decane, or + dodecane, or + tetradecane over the temperature range (288.15-308.15) K and atmospheric pressure. At this end, a densitometer Anton-Paar DMA 602 and a Ubbelohde viscosimeter were used. Excess molar volumes ($V_m^E$) are large and negative and decrease when the temperature is increased, which reveals that the main contribution to $V_m^E$ arises from structural effects. The values of the deviations of dynamic viscosity from linear dependence on mole fraction are also large and negative, indicating that $n$-alkanes are good breakers of the interactions between 1-iodonaphthalene molecules. Different models were applied for describing viscosity data. McAllister's equation correlates well with kinematic viscosities. Results are similar when dynamic viscosities ($\eta$) are correlated with the Grunberg-Nissan or Fang-He equations. This means that size effects are not relevant to the mentioned data. The adjustable parameter of the Grunberg-Nissan equation is negative for all the systems at any temperature, a typical feature of systems where dispersive interactions are dominant. This is in agreement with findings obtained in previous studies on similar $n$-alkane mixtures involving $C_6H_5X$ (X = Cl, Br, I) or 1,2,4-trichlorobenzene or 1-chloronaphthalene. Free volume effects have little influence on the present $\eta$ results, well represented by the absolute rate model using residual molar Gibbs energies obtained from the DISQUAC model.

## 1. Introduction

In order to attain a deeper understanding of the Patterson´s and Wilhelm´s effects [1-3], we are engaged in a systematic study of *n*-alkane mixtures containing a cyclic molecule. Particularly, we have considered systems with $C_6H_5X$ (X = F, Cl, Br, I) [4-7], or 1,2,4-trichlorobenzene, or 1-chloronaphthalene [5] or a bicyclic compound (decalin, tetralin…) [8]. The study is conducted by means of thermophysical properties, including excess molar properties at constant volume. In fact, we have demonstrated that the excess molar internal energy at constant volume is a key property to investigate the solutions cited above [5,8]. The systems have also been investigated using different models such as DISQUAC [9], Flory [10], or the concentration-concentration structure factor formalism [11]. As continuation of these works, volumetric and viscosity data for the mixtures 1-iodonaphthalene + heptane, or + decane, or + dodecane, or + tetradecane over the temperature range (288.15-308.15) K and at atmospheric pressure are now provided. These measurements are used to determine excess molar volumes ($V_m^E$) and molar quantities of activation of viscous flow: Gibbs energy ($\Delta G_m^*$), enthalpy($\Delta H_m^*$), and entropy ($\Delta S_m^*$) from the application of the Eyring's theory [12-14]. In addition, the new viscosity data are correlated by means of different equations: McAllister [15], Grunberg-Nissan [16], Fang-He [17], and using the model developed by Bloomfield and Dewan [18]. A survey of literature using the Thermolit NIST application shows that no data is available in the literature on these systems. Nevertheless, calorimetric [19], volumetric data [19] or viscosity data [20] are available for similar solutions with 1-chloronaphthalene.

Halogenated naphthalenes have applications in pharmaceuticals, dyes or as precursors of synthetic polymers. However, the replacement of a carbon-bond hydrogen by a halogen increases the inertness of the resulting molecule and hence its persistence in the environment, making these compounds organic pollutants [21]. In this framework, the knowledge of their physicochemical properties (e.g., hydrophobicity, polarity or volatility) is useful to investigate whether the compound prefers to dissolve in nonpolar, organic or polar environments. Iodinated mono-naphthalenes are less volatile and less toxic when compared with other mono-halogenated compounds [22]. 1-Iodonaphthalene has several interesting properties [23,24]: reactivity towards nucleophiles, low toxicity, precursor of several synthetic transformations, or easy attachment to organic compounds which has made it part of X-ray contrast materials in modern medicine.

## 2. Experimental

Pure liquids were used as received from the supplier. Information on the source and purity of the chemicals is included in Table 1. Solutions were prepared by weighing small vessels of about 10 cm³. The concentration of the mixtures (given by the mole fraction of 1-iodonaphthalene, $x_1$) was calculated from mass measurements. The masses were determined by weighing them using an analytical balance (MSU125P, Sartorius) and correcting for buoyancy effects, with a standard uncertainty of $5 \cdot 10^{-5}$ g. Along the process, caution was taken in order to prevent evaporation. Conversion to molar quantities was based on the relative atomic mass Table of 2015 issued by I.U.P.A.C [25]. The error in the final mole fraction is estimated to be 0.0010. All measurements were performed at atmospheric pressure. In addition, Pt-100 resistances, calibrated using the triple point of water and the melting point of Ga, were used to measure the temperature of the samples, with a standard uncertainty of 0.01 K.

**Table 1.** Sample description.

| Chemical name | CAS Number | Source | Initial purity[a] |
|---|---|---|---|
| 1-iodonaphthalene | 90-14-2 | Sigma-Aldrich | 0.988 |
| heptane | 142-82-5 | Sigma-Aldrich | 0.998 |
| decane | 124-18-5 | Sigma-Aldrich | 0.995 |
| dodecane | 112-40-3 | Sigma-Aldrich | 0.998 |
| tetradecane | 629-59-4 | Fluka | 0.995 |

[a]In mole fraction. Initial purity, measured by gas chromatography, certified by the supplier.

### 2.1 Density measurements

Densities ($\rho$) of the liquid samples were measured by means of a densitometer Anton Paar DMA 602. The temperature stability of the equipment was 0.01 K. The accuracy of the density measurements is estimated to be $5 \cdot 10^{-2}$ kg m$^{-3}$. The densities of the pure components were measured over time in order to test their stability, remaining constant within the uncertainty of the measurements. Further details on the calibration and on the applied procedure have been previously reported [26-28]. Nevertheless, it is worth mentioning that, in addition to the common pure liquids used in the calibration of the apparatus (heptane, cyclohexane, isooctane, toluene, and water) due to the high density of 1-iodonaphthalene, two additional liquids were also used: 2,4-dichlorotoluene and tetrachloroethylene. Table 2 compares our $\rho$ values for pure compounds over the temperature range (288.15-308.15) K with results from literature. Differences are lower than 0.1 %, which can be considered as the relative uncertainty for density. Our values of the isobaric expansion coefficients $\alpha_p$ (relative uncertainty 0.5 %) are also in good agreement with literature values (Table 2). For $V_m^E$, the uncertainty is estimated to be $0.011 \left| V_m^E \right|_{max} + 0.005$ cm$^3$ mol$^{-1}$ ($\left| V_{max}^E \right|$ denotes the maximum experimental value of the excess molar volume with respect to the mole fraction). For the excess isobaric expansion coefficient ($\alpha_p^E$), the corresponding relative uncertainty is 2 %.

**Table 2.** Densities ($\rho$), isobaric expansion coefficients ($\alpha_p$) and dynamic ($\eta$) viscosities of pure compounds at temperature $T$ and atmospheric pressure. Comparison of experimental (Exp.) results with literature (Lit.) values[a].

| $T$/K | 1-iodonaphthalene | | heptane | | decane | | dodecane | | tetradecane | |
|---|---|---|---|---|---|---|---|---|---|---|
| | Exp. | Lit. | Exp. | Lit. | Exp. | Lit. | Exp. | Lit. | Exp. | Lit. |
| $\rho$ /g cm⁻³ | | | | | | | | | | |
| 288.15 | 1.74524 | 1.74634[50] | 0.68797 | 0.68802[7] 0.68789[51] 0.68797[53] | 0.73367 | 0.73384[7] 0.7337[52] 0.73357[51] | 0.75255 | 0.75257[7] 0.7524[52] 0.76686[55] | 0.76631 | 0.76634[7] 0.7670[52] 0.76686[55] |
| 293.15 | 1.73959 | 1.7396[50] | 0.68378 | 0.68382[7] 0.68396[56] 0.6841[57] 0.68375[53] | 0.72990 | 0.73009[7] 0.7299[52] 0.72991[56] 0.72987[58] 0.73015[47] | 0.74893 | 0.74896[7] 0.7488[52] 0.74866[54] 0.74878[56] 0.74879[58] | 0.76277 | 0.76281[7] 0.7635[52] 0.76333[55] 0.76271[58] |
| 298.15 | 1.73394 | 1.73510[50] | 0.67954 | 0.67960[7,59] 0.67951[51] 0.67952[60] 0.67977[56] 0.6795[57] 0.67951[53] | 0.72612 | 0.72631[7] 0.7262[52] 0.72607[51] 0.72616[56] 0.72636[47] | 0.74529 | 0.74534[7] 0.7453[52] 0.74540[60] 0.74503[54] 0.74514[56] | 0.75923 | 0.75929[7] 0.7598[52] 0.75935[60] 0.75980[55] |
| 303.15 | 1.72828 | 1.72948[50] | 0.67529 | 0.67534[7] 0.67550[56] 0.6750[57] 0.67525[53] | 0.72234 | 0.72253[7] 0.7225[52] 0.72236[56] 0.72231[58] 0.72247[47] | 0.74167 | 0.74172[7] 0.7417[52] 0.74138[54] 0.74152[56] 0.74154[58] | 0.75570 | 0.75576[7] 0.7563[52] 0.75630[55] 0.75565[58] |
| 308.15 | 1.72265 | 1.72386[50] | 0.67100 | 0.67106[7] 0.67088[51] 0.6714[61] 0.6709[57] 0.67096[53] | 0.71854 | 0.71874[7] 0.7187[52] 0.71843[51] | 0.73803 | 0.73809[7] 0.7381[52] 0.73775[54] | 0.75217 | 0.75223[7] 0.7527[52] 0.7521[62] 0.74991[55] |
| $\alpha_p$ /10⁻³ K⁻¹ | | | | | | | | | | |
| 298.15 | 0.652 | 0.648[50] | 1.249 | 1.248[7] 1.254[53] 1.251[66] | 1.042 | 1.040[7] 1.043[64] | 0.974 | 0.971[7,63] 0.958[64] | 0.931 | 0.929[7] 0.938[65] 0.930[67] |
| $\eta$ /mPa s | | | | | | | | | | |
| 288.15 | 11.117 | | 0.437 | 0.433[53] | 0.998 | 0.991[68] 1.003[53] | 1.649 | 1.636[64] 1.635[54] | 2.607 | 2.596[64] 2.589[69] |
| 293.15 | 9.275 | | 0.413 | 0.410[53,46,70] 0.411[71] 0.413[56] | 0.922 | 0.916[47,70] 0.918[71] 0.922[58] | 1.499 | 1.485[47] 1.487[70] 1.490[72] 1.496[54] | 2.329 | 2.300[58] 2.330[55] 2.310[70] |
| 298.15 | 7.864 | | 0.391 | 0.390[53] 0.389[70,71] 0.392[75] 0.388[77] 0.393[20] | 0.855 | 0.849[64] 0.851[68,71] 0.848[47] 0.861[78] 0.859[79] 0.850[70] | 1.368 | 1.356[47] 1.360[71] 1.364[54] 1.367[76] 1.358[70] | 2.094 | 2.101[73] 2.085[754] 2.091[76] 2.087[69] 2.078[70] |
| 303.15 | 6.757 | | 0.371 | 0.371[53] 0.373[75] 0.368[77] 0.369[70] | 0.796 | 0.792[47] 0.794[58] 0.800[80] 0.791[70] | 1.254 | 1.248[47] 1.240[58] 1.250[72] 1.254[54] 1.246[70] | 1.894 | 1.895[73] 1.900[55] 1.882[70] |
| 308.15 | 5.873 | | 0.353 | 0.357[53] 0.351[70,71] 0.354[75] 0.350[61,77] | 0.743 | 0.743[53] 0.741[71] 0.739[70] | 1.155 | 1.149[64,70] 1.150[71] 1.157[54] | 1.722 | 1.715[73] 1.709[64] 1.717[69] |

[a]The uncertainties are: $u(T)$ = 0.02 K; $u(p)$ = 10 kPa. The relative combined expanded uncertainties (0.95 level of confidence) are: $U_{rc}(\rho)$ = 0.002; $U_{rc}(\alpha_p)$ = 0.01, and $U_{rc}(\eta)$ = 0.022 .

Kinematic viscosities ($\nu$) were determined by means of an Ubbelohde viscosimeter with a Schott-Geräte automatic measuring unit model AVS-350. The temperature was held constant within $\pm 0.02$ K using a controller bath CT52 (from Schott). Details on the calibration of the apparatus can be found elsewhere [29]. Values of dynamic viscosity ($\eta = \nu \rho$) were obtained using the densities determined in the present work. The uncertainties of the kinematic and dynamic viscosities are estimated at 1% and 1.1 %, respectively [29]. Table 2 shows a rather good agreement exists between the present values of $\eta$ for the pure compounds with literature results (see Table S1, supplementary material, for a similar comparison for $\nu$). The mean deviations for the $\eta$ values of *n*-alkanes at any temperature are lower than 1.1 %. For kinematic viscosities, the mean deviations are similar. For 1-iodonaphthalene, no viscosity data have encountered in the open literature for comparison. The relative uncertainty of the deviations of dynamic viscosity from linear dependence on mole fraction (hereafter, deviations in absolute viscosity, see below) does not exceed from 2.5%. This value is supported by the comparison of our results with those from the literature for the 1-propanol + dimethyl carbonate mixture at (288.15-313.15) K [30].

## 3. Models

### 3.1 Absolute reaction rate model

This theory relates viscosity to the free energy required for a molecule to flow from an equilibrium position to a new one, overcoming the attractive interactions caused by its neighbors [12-14]. The expression for dynamic viscosity is:

$$\eta = \frac{hN_{\mathrm{A}}}{V_{\mathrm{m}}}\exp\left(\frac{\Delta G_{\mathrm{m}}^{*}}{RT}\right) \tag{1}$$

where $h$ is the Planck's constant, $N_{\mathrm{A}}$ is the Avogadro's number and $V_{\mathrm{m}}$ is the molar volume. The values of $\Delta G_{\mathrm{m}}^{*}(=\Delta H_{\mathrm{m}}^{*} - T\Delta S_{\mathrm{m}}^{*})$ were calculated determining previously the required values of $\Delta H_{\mathrm{m}}^{*}$ and $\Delta S_{\mathrm{m}}^{*}$ from the plots $\ln\left(\dfrac{\eta V_{\mathrm{m}}}{hN_{\mathrm{A}}}\right)$ *vs.* $1/T$ [31,32]. These ones give a straight line for each mixture and $\Delta H_{\mathrm{m}}^{*}$, and $\Delta S_{\mathrm{m}}^{*}$ can be estimated from its slope and intercept.

### 3.2 Kinematic viscosity

Results of kinematic viscosities have been correlated using the McAllister equation, based on the Eyring's theory [14] and a three-body interaction model [15]:

$$\ln \nu = x_1^3 \ln \nu_1 + 3x_1^2 x_2 \ln Z_{12} + 3x_1 x_2^2 \ln Z_{21} + x_2^3 \ln \nu_2 - \ln\left(x_1 + x_2 \frac{M_2}{M_1}\right)$$

$$+ 3x_1^2 x_2 \ln\left(\frac{2}{3} + \frac{M_2}{3M_1}\right) + 3x_1 x_2^2 \ln\left(\frac{1}{3} + \frac{2M_2}{3M_1}\right) + x_2^3 \ln\left(\frac{M_2}{M_1}\right) \tag{2}$$

where $M_i$ and $\nu_i$ are, respectively, the molar mass and the kinematic viscosity of component i, and $Z_{12}, Z_{21}$ the adjustable parameters.

### 3.3 Dynamic viscosity: equations with one adjustable parameter

Values of dynamic viscosity were correlated using the Grunberg-Nissan [16] and the Fang-He [17] equations. According to the former equation, viscosity is calculated from the expression [16]:

$$\ln \eta = x_1 \ln \eta_1 + x_2 \ln \eta_2 + x_1 x_2 G_{12} \tag{3}$$

where $\eta_i$ stands for the dynamic viscosity of component i. The Fang-He equation combines the Eyring's model with a modified Flory-Huggins equation expressed in terms of surface fractions [17]:

$$\ln \eta = \left(\theta_1 \ln \eta_1 + \theta_2 \ln \eta_2\right) + \left(x_1 \ln \frac{\theta_1}{x_1} + x_2 \ln \frac{\theta_2}{x_2}\right) + \left(x_1 V_{m1}^{2/3} + x_2 V_{m2}^{2/3}\right)\left(\frac{W_{12}}{RT}\right)\theta_1 \theta_2 \tag{4}$$

where $\theta_i$ is the molecular surface fraction defined by $\theta_i = \dfrac{x_i V_{mi}^{2/3}}{\sum x_j V_{mj}^{2/3}}$, being $V_{mi}$ the molar volume of component i). In equations (3) and (4), the adjustable parameters are, respectively, $G_{12}$ and $W_{12}$.

### 3.4 Bloomfield-Dewan's model

This theory [18] combines the absolute reaction rate model with the free volume theory [33]. The latter relates viscosity to the probability of occurrence of an empty neighboring site where a molecule can jump. Thus, dynamic viscosity can be determined from the equation:

$$\ln \eta = \left(x_1 \ln \eta_1 + x_2 \ln \eta_2\right) + \alpha \ln \eta_{fv} + \beta \ln \eta_{ar} \tag{5}$$

This means that the probability for viscous flow is calculated as the product of the probabilities of having sufficient activation energy and of the existence of an empty site [18,34]. The parameters $\alpha, \beta$ are weighting factors with values between 0 and 1. In equation (5), $\ln \eta_{fv}$ arises from free volume effects. In the present application, calculations have been conducted assuming that $\alpha = 0$ (see below). The contribution arising from the absolute reaction rate model is given by:

$$\ln \eta_{\mathrm{ar}} = -\frac{\Delta G_{\mathrm{m}}^{\mathrm{RES}}}{RT} \qquad (6)$$

Typically, the values of $\Delta G_{\mathrm{m}}^{\mathrm{RES}}$ are obtained from the application of the Flory model using the corresponding expressions for $H_{\mathrm{m}}^{\mathrm{E}}$ and for the residual entropy [18]. Here, $\Delta G_{\mathrm{m}}^{\mathrm{RES}}$ values have been calculated using the DISQUAC model. Essentially, $\Delta G_{\mathrm{m}}^{\mathrm{RES}}$ is obtained from the excess Gibbs energy, $G_{\mathrm{m}}^{\mathrm{E}}$, by subtracting the corresponding combinatorial term represented by the Flory-Huggins equation. The interaction parameters needed were obtained from a detailed investigation of $C_6H_5X$ (X = F, Cl, Br, I) or 1-chloronaphthalene + $n$-alkane mixtures using DISQUAC. Particularly, for the (iodine/aliphatic) contacts, the Gibbs energy interchange coefficients, needed for calculations at 298.15 K, are: $-1.15$ (dispersive coefficient) and 2.0 (quasichemical coefficient). For the (iodine/aromatic) contacts, such interchange coefficients are assumed to be 0.

## 4. Results

### 4.1 Experimental results

Table S2 contains values of density and of kinematic viscosity for the studied systems. Table 3 lists results on $V_{\mathrm{m}}^{\mathrm{E}}$, $\eta$ and $\Delta\eta$, the deviations in absolute viscosity (a non-Gibbsian property), which were determined from the equation:

$$\Delta\eta = \eta - (x_1\eta_1 + x_2\eta_2) \qquad (7)$$

**Table 3.** Excess molar volumes ($V_{\mathrm{m}}^{\mathrm{E}}$) and dynamic viscosities ($\eta$) of 1-iodonaphthalene (1) + $n$-alkane (2) mixtures at temperature $T$ and atmospheric pressure. Values of $\Delta\eta$ (Eq. 7) are also given[a].

| $x_1$ | $V_{\mathrm{m}}^{\mathrm{E}}$ /cm³ mol⁻¹ | $\eta$ /mPa s | $\Delta\eta$ /mPa s | $V_{\mathrm{m}}^{\mathrm{E}}$ /cm³ mol⁻¹ | $\eta$ /mPa s | $\Delta\eta$ /mPa s |
|---|---|---|---|---|---|---|
| | | | 1-iodonaphthalene (1) + heptane (2) | | | |
| | $T$/K = 288.15 | | | $T$/K = 293.15 | | |
| 0.0956 | $-0.609$ | 0.538 | $-0.920$ | $-0.632$ | 0.507 | $-0.753$ |
| 0.1976 | $-1.037$ | 0.683 | $-1.838$ | $-1.081$ | 0.642 | $-1.521$ |
| 0.2993 | $-1.280$ | 0.894 | $-2.740$ | $-1.334$ | 0.832 | $-2.233$ |
| 0.3980 | $-1.368$ | 1.177 | $-3.511$ | $-1.426$ | 1.086 | $-2.854$ |
| 0.4989 | $-1.405$ | 1.581 | $-4.184$ | $-1.461$ | 1.448 | $-3.386$ |
| 0.5993 | $-1.320$ | 2.164 | $-4.674$ | $-1.369$ | 1.960 | $-3.765$ |
| 0.6970 | $-1.126$ | 3.014 | $-4.867$ | $-1.170$ | 2.699 | $-3.891$ |
| 0.7982 | $-0.874$ | 4.408 | $-4.553$ | $-0.905$ | 3.876 | $-3.611$ |
| 0.8924 | $-0.554$ | 6.584 | $-3.383$ | $-0.572$ | 5.674 | $-2.647$ |
| | $T$/K = 298.15 | | | $T$/K = 303.15 | | |
| 0.0956 | $-0.660$ | 0.479 | $-0.627$ | $-0.690$ | 0.453 | $-0.528$ |
| 0.1976 | $-1.136$ | 0.605 | $-1.263$ | $-1.199$ | 0.570 | $-1.063$ |
| 0.2993 | −1.403 | 0.778 | −1.850 | −1.483 | 0.729 | −1.554 |
| 0.3980 | −1.502 | 1.011 | −2.355 | −1.591 | 0.941 | −1.972 |
| 0.4989 | −1.537 | 1.334 | −2.786 | −1.626 | 1.230 | −2.327 |
| 0.5993 | −1.439 | 1.790 | −3.079 | −1.522 | 1.643 | −2.555 |
| 0.6970 | −1.218 | 2.426 | −3.174 | −1.267 | 2.198 | −2.625 |
| 0.7982 | −0.940 | 3.440 | −2.917 | −0.976 | 3.073 | −2.396 |
| 0.8924 | −0.592 | 4.962 | −2.097 | −0.612 | 4.371 | −1.698 |

$T/K = 308.15$

| | | | |
|---|---|---|---|
| 0.0956 | −0.724 | 0.429 | −0.451 |
| 0.1976 | −1.244 | 0.538 | −0.906 |
| 0.2993 | −1.538 | 0.684 | −1.321 |
| 0.3980 | −1.650 | 0.877 | −1.672 |
| 0.4989 | −1.682 | 1.141 | −1.966 |
| 0.5993 | −1.568 | 1.508 | −2.153 |
| 0.6970 | −1.321 | 2.002 | −2.199 |
| 0.7982 | −1.014 | 2.767 | −1.992 |
| 0.8924 | −0.634 | 3.889 | −1.390 |

1-iodonaphthalene (1) + decane (2)

| $T/K = 288.15$ | | | | $T/K = 293.15$ | | |
|---|---|---|---|---|---|---|
| 0.0981 | −0.317 | 1.130 | −0.866 | −0.328 | 1.042 | −0.703 |
| 0.1958 | −0.473 | 1.299 | −1.690 | −0.492 | 1.194 | −1.369 |
| 0.2960 | −0.629 | 1.529 | −2.479 | −0.655 | 1.398 | −2.005 |
| 0.3971 | −0.720 | 1.840 | −3.196 | −0.750 | 1.673 | −2.577 |
| 0.4981 | −0.762 | 2.266 | −3.797 | −0.792 | 2.047 | −3.049 |
| 0.5962 | −0.747 | 2.843 | −4.218 | −0.776 | 2.543 | −3.376 |
| 0.6957 | −0.678 | 3.666 | −4.406 | −0.703 | 3.242 | −3.510 |
| 0.7968 | −0.538 | 4.948 | −4.152 | −0.557 | 4.330 | −3.270 |
| 0.8958 | −0.341 | 7.005 | −3.102 | −0.352 | 6.005 | −2.424 |

| $T/K = 298.15$ | | | | $T/K = 303.15$ | | |
|---|---|---|---|---|---|---|
| 0.0981 | −0.339 | 0.964 | −0.580 | −0.351 | 0.895 | −0.485 |
| 0.1958 | −0.511 | 1.101 | −1.128 | −0.532 | 1.020 | −0.941 |
| 0.2960 | −0.682 | 1.284 | −1.649 | −0.709 | 1.185 | −1.374 |
| 0.3971 | −0.780 | 1.529 | −2.113 | −0.812 | 1.405 | −1.755 |
| 0.4981 | −0.824 | 1.858 | −2.493 | −0.857 | 1.695 | −2.066 |
| 0.5962 | −0.806 | 2.291 | −2.749 | −0.837 | 2.076 | −2.270 |
| 0.6957 | −0.730 | 2.900 | −2.838 | −0.757 | 2.615 | −2.324 |
| 0.7968 | −0.577 | 3.816 | −2.632 | −0.598 | 3.394 | −2.147 |
| 0.8958 | −0.364 | 5.216 | −1.926 | −0.376 | 4.583 | −1.547 |

$T/K = 308.15$

| | | | |
|---|---|---|---|
| 0.0981 | −0.364 | 0.834 | −0.412 |
| 0.1958 | −0.554 | 0.948 | −0.799 |
| 0.2960 | −0.740 | 1.098 | −1.162 |
| 0.3971 | −0.846 | 1.296 | −1.482 |
| 0.4981 | −0.892 | 1.556 | −1.740 |
| 0.5962 | −0.870 | 1.893 | −1.906 |
| 0.6957 | −0.786 | 2.362 | −1.947 |
| 0.7968 | −0.621 | 3.054 | −1.773 |
| 0.8958 | −0.388 | 4.055 | −1.279 |

## 1-iodonaphthalene (1) + dodecane (2)

| | T/K =288.15 | | | T/K =293.15 | |
|---|---|---|---|---|---|
| 0.0996 | −0.177 | 1.790 | −0.802 | −0.184 | 1.622 | −0.651 |
| 0.1981 | −0.327 | 1.969 | −1.557 | −0.339 | 1.781 | −1.258 |
| 0.2994 | −0.456 | 2.207 | −2.277 | −0.473 | 1.990 | −1.838 |
| 0.3981 | −0.540 | 2.519 | −2.900 | −0.559 | 2.259 | −2.336 |
| 0.4971 | −0.582 | 2.927 | −3.429 | −0.604 | 2.616 | −2.749 |
| 0.5974 | −0.574 | 3.509 | −3.796 | −0.595 | 3.107 | −3.038 |
| 0.6975 | −0.542 | 4.304 | −3.948 | −0.562 | 3.784 | −3.139 |
| 0.7987 | −0.426 | 5.487 | −3.725 | −0.441 | 4.773 | −2.938 |
| 0.8990 | −0.263 | 7.407 | −2.753 | −0.271 | 6.343 | −2.147 |

| | T/K = 298.15 | | | T/K =303.15 | |
|---|---|---|---|---|---|
| 0.0996 | −0.190 | 1.480 | −0.535 | −0.196 | 1.357 | −0.445 |
| 0.1981 | −0.351 | 1.620 | −1.035 | −0.364 | 1.483 | −0.861 |
| 0.2994 | −0.490 | 1.806 | −1.507 | −0.506 | 1.646 | −1.256 |
| 0.3981 | −0.580 | 2.044 | −1.910 | −0.602 | 1.855 | −1.590 |
| 0.4971 | −0.626 | 2.353 | −2.245 | −0.649 | 2.132 | −1.858 |
| 0.5974 | −0.617 | 2.778 | −2.471 | −0.640 | 2.499 | −2.042 |
| 0.6975 | −0.582 | 3.365 | −2.534 | −0.602 | 3.008 | −2.085 |
| 0.7987 | −0.456 | 4.194 | −2.363 | −0.472 | 3.713 | −1.937 |
| 0.8990 | −0.280 | 5.495 | −1.713 | −0.288 | 4.799 | −1.403 |

| | T/K = 308.15 | | |
|---|---|---|---|
| 0.0996 | −0.203 | 1.246 | −0.378 |
| 0.1981 | −0.375 | 1.363 | −0.727 |
| 0.2994 | −0.526 | 1.509 | −1.058 |
| 0.3981 | −0.624 | 1.696 | −1.338 |
| 0.4971 | −0.674 | 1.936 | −1.564 |
| 0.5974 | −0.666 | 2.263 | −1.710 |
| 0.6975 | −0.624 | 2.705 | −1.741 |
| 0.7987 | −0.487 | 3.313 | −1.611 |
| 0.8990 | −0.297 | 4.247 | −1.150 |

## 1-iodonaphthalene (1) + tetradecane (2)

| | T/K =288.15 | | | T/K = 293.15 | |
|---|---|---|---|---|---|
| 0.0988 | −0.131 | 2.750 | −0.698 | −0.135 | 2.454 | −0.561 |
| 0.1996 | −0.243 | 2.915 | −1.391 | −0.252 | 2.598 | −1.118 |
| 0.2992 | −0.334 | 3.129 | −2.025 | −0.347 | 2.792 | −1.615 |
| 0.3962 | −0.404 | 3.413 | −2.566 | −0.420 | 3.015 | −2.067 |
| 0.4971 | −0.435 | 3.799 | −3.039 | −0.452 | 3.350 | −2.432 |
| 0.5968 | −0.456 | 4.311 | −3.375 | −0.473 | 3.796 | −2.679 |
| 0.6973 | −0.435 | 5.064 | −3.476 | −0.451 | 4.422 | −2.751 |
| 0.7984 | −0.358 | 6.149 | −3.252 | −0.370 | 5.349 | −2.526 |
| 0.8983 | −0.239 | 7.853 | −2.399 | −0.246 | 6.689 | −1.881 |

| | T/K =298.15 | | | T/K =303.15 | |
|---|---|---|---|---|---|
| 0.0988 | −0.140 | 2.201 | −0.463 | −0.144 | 1.986 | −0.388 |
| 0.1996 | −0.261 | 2.328 | −0.918 | −0.270 | 2.102 | −0.763 |
| 0.2992 | −0.359 | 2.500 | −1.321 | −0.372 | 2.250 | −1.099 |
| 0.3962 | −0.435 | 2.715 | −1.665 | −0.451 | 2.432 | −1.389 |
| 0.4971 | −0.470 | 2.996 | −1.966 | −0.487 | 2.676 | −1.635 |
| 0.5968 | −0.490 | 3.362 | −2.176 | −0.507 | 2.997 | −1.799 |
| 0.6973 | −0.467 | 3.893 | −2.225 | −0.483 | 3.446 | −1.839 |
| 0.7984 | −0.383 | 4.669 | −2.032 | −0.395 | 4.089 | −1.688 |
| 0.8983 | −0.253 | 5.792 | −1.486 | −0.260 | 5.053 | −1.210 |
| $T/K$ =308.15 | | | | | | |
| 0.0988 | −0.148 | 1.805 | −0.328 | | | |
| 0.1996 | −0.278 | 1.904 | −0.647 | | | |
| 0.2992 | −0.384 | 2.045 | −0.920 | | | |
| 0.3962 | −0.466 | 2.193 | −1.174 | | | |
| 0.4971 | −0.503 | 2.417 | −1.369 | | | |
| 0.5968 | −0.524 | 2.697 | −1.502 | | | |
| 0.6973 | −0.499 | 3.083 | −1.534 | | | |
| 0.7984 | −0.407 | 3.640 | −1.397 | | | |
| 0.8983 | −0.268 | 4.471 | −0.981 | | | |

[a]The uncertainties, $u$, are: $u(T)$ = 0.02 K; $u(p)$ = 10 kPa; $u(x_1)$ = 0.0010; $u(V_m^E)$ = 0.011 $\left| V_m^E \right|_{max}$ + 0.005 cm$^3$ mol$^{-1}$. The relative combined expanded uncertainties (0.95 level of confidence) for $\eta$ and $\Delta\eta$ are, respectively, $U_{rc}(\eta)$ = 0.022 and $U_{rc}(\Delta\eta)$ = 0.05.

For a given mixture, the value of $\alpha_p$ at 298.15 K at each concentration was obtained from linear regressions of density results *vs.* temperature. Values of $\alpha_p^E$ (Table S3) were determined from the equation:

$$\alpha_p^E = \alpha_p - \phi_1\alpha_{p1} - \phi_2\alpha_{p2} \tag{8}$$

where $\phi_i = x_1 V_{mi} / (V_{m1} + x_2 V_{m2})$ is the volume fraction of component i. Results at 298.15 K for $V_m^E$, $\eta$, $\Delta\eta$, and $\alpha_p^E$ are shown graphically in Figures 1 to 4. Values of $\Delta G_m^*$, $\Delta H_m^*$, $\Delta S_m^*$ and of $\Delta(\Delta G_m^*) = \Delta G_m^* - x_1\Delta G_{m1}^* - x_2\Delta G_{m2}^*$ are collected in Table S4 of supplementary material (see Figure 5). On the other hand, results on $Q = V_m^E, \alpha_p^E$ and $\Delta(\Delta G_m^*)$ data were fitted to the equation:

$$Q = x_1(1-x_1)\sum_{i=0}^{k-1} A_i(2x_1-1)^i \tag{9}$$

The number of coefficients $k$ used in equation (9) for each mixture was determined by applying an F-test [35] at the 99.5 % confidence level. Finally, values of $\Delta\eta$ were fitted by unweighted least-squares polynomial regression to the equation:

$$\Delta\eta = x_1(1-x_1)\left( A_0 + \frac{A_1 x_1}{x_1 + A_2(1-x_1)} \right) \tag{10}$$

Table 4 lists the parameters $A_i$, obtained in the corresponding regressions to the equations (9) and (10), together with the standard deviations, $\sigma(R)$, defined by:

$$\sigma(R) = \left[ \frac{1}{N-k} \sum \left( R_{\text{cal}} - R_{\text{exp}} \right)^2 \right]^{1/2} \qquad (11)$$

where $N$ is the number of direct experimental values and $R = Q, \Delta\eta$.

### 4.2 Results from models.

For viscosity, results obtained from the application of the different models/equations considered in the present study are collected in Tables 5 and Table S5, which also list the adjusted parameters: $Z_{12}, Z_{21}$ (equation (2), Table 5), $G_{12}$ (equation (3), Tables 5 and S5), $W_{12}/RT$ (equation (4), Table S5), as well as the values of $\alpha$ and $\beta$ (equation (5), Table 5, Figure 2). Particularly, results of the correlations are compared by means of the relative standard deviations, $\sigma_r(F)$, calculated from:

$$\sigma_r(F = \nu, \eta) = \left[ \frac{1}{N} \sum \left( \frac{F_{\text{cal}} - F_{\text{exp}}}{F_{\text{exp}}} \right)^2 \right]^{1/2} \qquad (12)$$

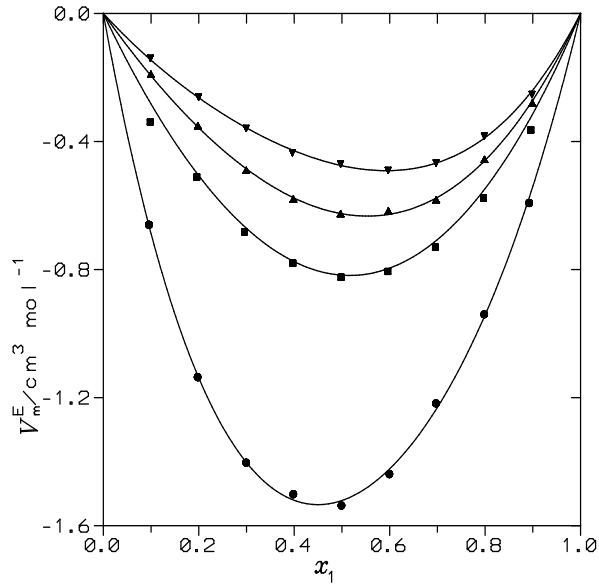

**Figure 1.** $V_m^E$ of 1-iodonaphthalene (1) + $n$-alkane (2) mixtures at 298.15 K and atmospheric pressure. Points, experimental results (this work): (●), heptane; (■), decane; (▲), dodecane; (▼), tetradecane. Solid lines, calculations using coefficients listed in Table 4.

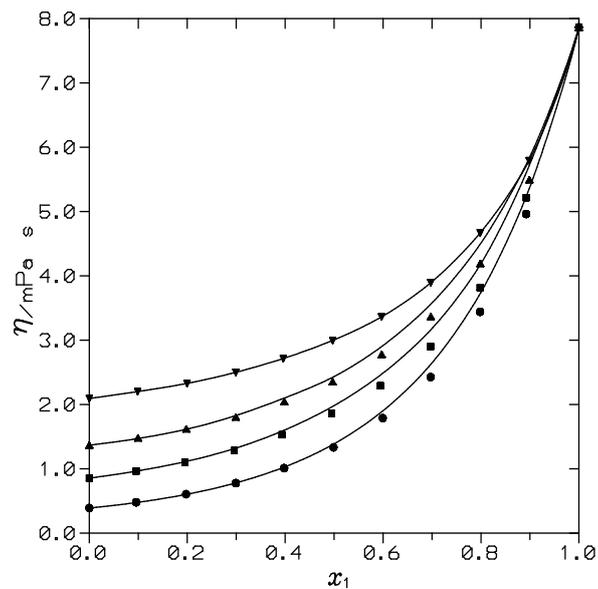

**Figure 2.** Dynamic viscosities ($\eta$) of 1-iodonaphthalene (1) + $n$-alkane (2) mixtures at 298.15 K and atmospheric pressure. Points, experimental results (this work): (●), heptane; (■), decane; (▲), dodecane; (▼), tetradecane. Solid lines, results from the application of the Bloomfield-Dewan´s model (equation 5) with ($\alpha = 0$, $\beta = 1$).

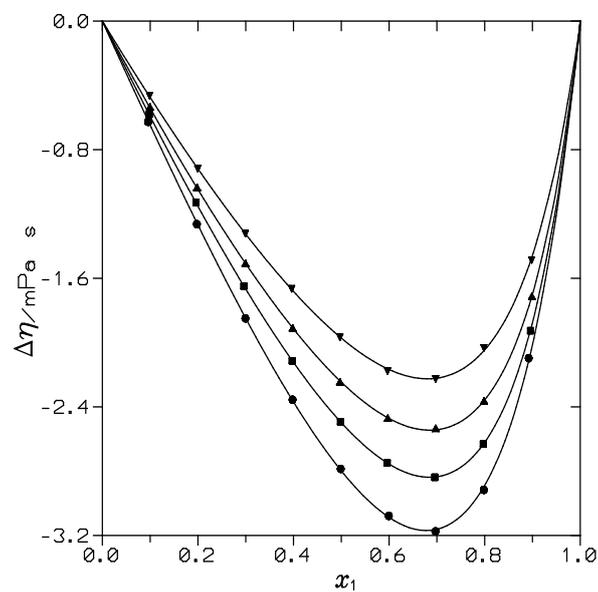

**Figure 3.** $\Delta\eta$ of 1-iodonaphthalene (1) + $n$-alkane (2) mixtures at 298.15 K and atmospheric pressure. Points, experimental results (this work): (●), heptane; (■), decane; (▲), dodecane; (▼), tetradecane. Solid lines, calculations using coefficients listed in Table 4.

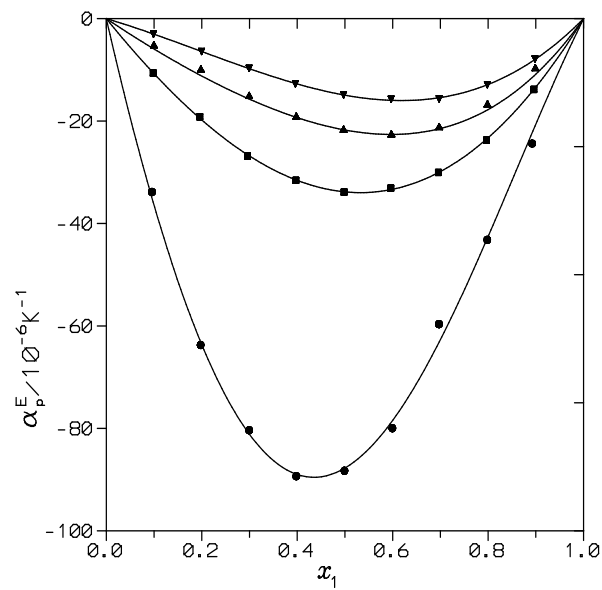

**Figure 4.** $\alpha_p^E$ of 1-iodonaphthalene (1) + *n*-alkane (2) mixtures at 298.15 K and atmospheric pressure. Points, experimental results (this work): (●), heptane; (■), decane; (▲), dodecane; (▼), tetradecane. Solid lines, calculations using coefficients listed in Table 4.

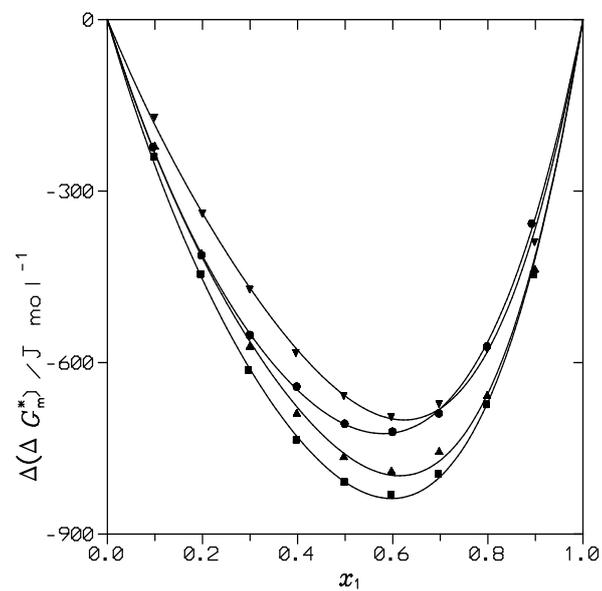

**Figure 5.** $\Delta(\Delta G_m^*)$ of 1-iodonaphthalene (1) + *n*-alkane (2) mixtures at 298.15 K and atmospheric pressure. Points, experimental results (this work): (●), heptane; (■), decane; (▲), dodecane; (▼), tetradecane. Solid lines, calculations using coefficients listed in Table 4.

**Table 4.** Coefficients $A_i$ and standard deviations, $\sigma(R)$ (eq. (11)) for representation of the $R^a$ property at temperature $T$ and atmospheric pressure for 1-iodonaphthalene (1) + $n$-alkane (2) systems by eqs. (9) and (10) (between parentheses, standard deviations of the coefficients are given).

| Property | $T$/K | $A_0$ | $A_1$ | $A_2$ | $\sigma^{\,b}$ |
|---|---|---|---|---|---|
| | | 1-iodonaphthalene (1) + heptane (2) | | | |
| $V_m^E$ /cm$^3$ mol$^{-1}$ | 288.15 | $-5.560\,(\pm0.033)$ | $0.857(\pm0.067)$ | $-1.20\,(\pm0.15)$ | 0.015 |
| $\Delta\eta$ / mPa s | 288.15 | $-9.63\,(\pm0.06)$ | $-39.9\,(\pm0.3)$ | $4.56\,(\pm0.08)$ | 0.011 |
| $V_m^E$ /cm$^3$ mol$^{-1}$ | 293.15 | $-5.785\,(\pm0.032)$ | $0.923\,(\pm0.067)$ | $-1.22\,(\pm0.15)$ | 0.014 |
| $\Delta\eta$ mPa s | 293.15 | $-7.91\,(\pm0.06)$ | $-29.9\,(\pm0.3)$ | $4.26\,(\pm0.08)$ | 0.009 |
| $V_m^E$ /cm$^3$ mol$^{-1}$ | 298.15 | $-6.086\,(\pm0.037)$ | $1.025\,(\pm0.074)$ | $-1.16(\pm0.17)$ | 0.016 |
| $\Delta\eta$ /mPa s | 298.15 | $-6.56\,(\pm0.06)$ | $-22.7\,(\pm0.3)$ | $3.90\,(\pm0.12)$ | 0.011 |
| $\alpha_p^E$ /10$^{-6}$ K$^{-1}$ | 298.15 | $-70.97\ 10^{-5}$ $(\pm0.12\ 10^{-5})$ | $-4.7\ 10^{-5}$ $(\pm0.2\ 10^{-5})$ | $-5.5\ 10^{-5}$ $(\pm0.5\ 10^{-5})$ | $5\ 10^{-7}$ |
| $\Delta\left(\Delta G_m^*\right)$ /J mol$^{-1}$ | 298.15 | $-2805\,(\pm19)$ | $-813\,(\pm38)$ | $-651\,(\pm87)$ | 8 |
| $V_m^E$ /cm$^3$ mol$^{-1}$ | 303.15 | $-6.439\,(\pm0.047)$ | $1.163\,(\pm0.094)$ | $-1.02\,(\pm0.22)$ | 0.015 |
| $\Delta\eta$ /mPa s | 303.15 | $-5.54\,(\pm0.06)$ | $-17.8\,(\pm0.3)$ | $3.67\,(\pm0.14)$ | 0.011 |
| $V_m^E$ /cm$^3$ mol$^{-1}$ | 308.15 | $-6.661\,(\pm0.039)$ | $1.216(\pm0.077)$ | $-1.15\,(\pm0.18)$ | 0.017 |
| $\Delta\eta$ /mPa s | 308.15 | $-4.72\,(\pm0.06)$ | $-14.0\,(\pm0.3)$ | $3.39\,(\pm0.15)$ | 0.011 |
| | | 1-iodonaphthalene (1) + decane (2) | | | |
| $V_m^E$ /cm$^3$ mol$^{-1}$ | 288.15 | $-3.020\,(\pm0.035)$ | $-0.214\,(\pm0.069)$ | $-0.66\,(\pm0.16)$ | 0.015 |
| $\Delta\eta$ /mPa s | 288.15 | $-9.021\,(\pm0.010)$ | $-39.16\,(\pm0.08)$ | $5.31\,(\pm0.02)$ | 0.002 |
| $V_m^E$ /cm$^3$ mol$^{-1}$ | 293.15 | $-3.142\,(\pm0.035)$ | $-0.209\,(\pm0.070)$ | $-0.65\,(\pm0.16)$ | 0.015 |
| $\Delta\eta$ /mPa s | 293.15 | $-7.35\,(\pm0.02)$ | $-29.44\,(\pm0.15)$ | $5.03\,(\pm0.05)$ | 0.004 |
| $V_m^E$ /cm$^3$ mol$^{-1}$ | 298.15 | $-3.270\,(\pm0.035)$ | $-0.209\,(\pm0.070)$ | $-0.65\,(\pm0.16)$ | 0.015 |
| $\Delta\eta$ /mPa s | 298.15 | $-6.077\,(\pm0.010)$ | $-22.60\,(\pm0.06)$ | $4.77\,(\pm0.03)$ | 0.0019 |
| $\alpha_p^E$ /10$^{-6}$ K$^{-1}$ | 298.15 | $-135.3\ 10^{-6}$ $(\pm0.4\ 10^{-6})$ | $-18.1\ 10^{-6}$ $(\pm0.9\ 10^{-6})$ | $2\ 10^{-6}$ $(\pm2\ 10^{-6})$ | $2\ 10^{-7}$ |
| $\Delta\left(\Delta G_m^*\right)$ /J mol$^{-1}$ | 298.15 | $-3237\,(\pm20)$ | $-1118\,(\pm40)$ | $-768\,(\pm91)$ | 9 |
| $V_m^E$ /cm$^3$ mol$^{-1}$ | 303.15 | $-3.400\,(\pm0.035)$ | $-0.205\,(\pm0.071)$ | $-0.65\,(\pm0.16)$ | 0.015 |
| $\Delta\eta$ /mPa s | 303.15 | $-5.085\,(\pm0.011)$ | $-17.44\,(\pm0.06)$ | $4.46\,(\pm0.03)$ | 0.002 |
| $V_m^E$ /cm$^3$ mol$^{-1}$ | 308.15 | $-3.540\,(\pm0.035)$ | $-0.198\,(\pm0.071)$ | $-0.65\,(\pm0.16)$ | 0.015 |
| $\Delta\eta$ /mPa s | 308.15 | $-4.329\,(\pm0.019)$ | $-13.97\,(\pm0.11)$ | $4.29\,(\pm0.07)$ | 0.004 |
| | | 1-iodonaphthalene (1) + dodecane (2) | | | |
| $V_m^E$ /cm$^3$ mol$^{-1}$ | 288.15 | $-2.327\,(\pm0.016)$ | $-0.497\,(\pm0.031)$ | $-0.145\,(\pm0.372)$ | 0.007 |
| $\Delta\eta$ /mPa s | 288.15 | $-8.32\,(\pm0.03)$ | $-35.9\,(\pm0.2)$ | $5.64\,(\pm0.07)$ | 0.005 |
| $V_m^E$ /cm$^3$ mol$^{-1}$ | 293.15 | $-1.879\,(\pm0.016)$ | $-0.512\,(\pm0.031)$ | $-0.142\,(\pm0.072)$ | 0.007 |
| $\Delta\eta$ /mPa s | 293.15 | $-6.754\,(\pm0.017)$ | $-26.98\,(\pm0.13)$ | $5.33\,(\pm0.05)$ | 0.004 |
| $V_m^E$ /cm$^3$ mol$^{-1}$ | 298.15 | $-2.503\,(\pm0.016)$ | $-0.528\,(\pm0.032)$ | $-0.131\,(\pm0.074)$ | 0.007 |
| $\Delta\eta$ /mPa s | 298.15 | $-5.58\,(\pm0.02)$ | $-20.94\,(\pm0.18)$ | $5.15\,(\pm0.09)$ | 0.005 |
| $\alpha_p^E$ /10$^{-6}$ K$^{-1}$ | 298.15 | $-87.0\ 10^{-6}$ $(\pm0.4\ 10^{-6})$ | $-34.1\ 10^{-6}$ $(\pm0.8\ 10^{-6})$ | $9.8\ 10^{-6}$ $(\pm1.8\ 10^{-6})$ | $1.7\ 10^{-7}$ |
| $\Delta\left(\Delta G_m^*\right)$ /J mol$^{-1}$ | 298.15 | $-3042\,(\pm28)$ | $-1234\,(\pm57)$ | $-851\,(\pm131)$ | 12 |
| $V_m^E$ /cm$^3$ mol$^{-1}$ | 303.15 | $-2.595\,(\pm0.015)$ | $-0.544\,(\pm0.031)$ | $-0.113\,(\pm0.071)$ | 0.007 |
| $\Delta\eta$ /mPa s | 303.15 | $-4.68\,(\pm0.03)$ | $-16.90\,(\pm0.19)$ | $5.13\,(\pm0.12)$ | 0.006 |
| $V_m^E$ /cm$^3$ mol$^{-1}$ | 308.15 | $-2.708\,(\pm0.012)$ | $-0.562\,(\pm0.031)$ | | 0.007 |
| $\Delta\eta$ /mPa s | 308.15 | $-3.95\,(\pm0.02)$ | $-13.41\,(\pm0.14)$ | $4.82\,(\pm0.11)$ | 0.004 |

|  |  | 1-iodonaphthalene (1) + tetradecane (2) |  |  |  |
|---|---|---|---|---|---|
| $V_m^E$ /cm$^3$ mol$^{-1}$ | 288.15 | $-1.765\ (\pm 0.016)$ | $-0.611\ (\pm 0.032)$ | $-0.383\ (\pm 0.074)$ | 0.007 |
| $\Delta\eta$ /mPa s | 288.15 | $-7.39\ (\pm 0.05)$ | $-29.9\ (\pm 0.3)$ | $5.24\ (\pm 0.12)$ | 0.010 |
| $V_m^E$ /cm$^3$ mol$^{-1}$ | 293.15 | $-1.834\ (\pm 0.017)$ | $-0.627\ (\pm 0.033)$ | $-0.372\ (\pm 0.077)$ | 0.007 |
| $\Delta\eta$ /mPa s | 293.15 | $-6.01\ (\pm 0.09)$ | $-22.6\ (\pm 0.6)$ | $5.1\ (\pm 0.3)$ | 0.018 |
| $V_m^E$ /cm$^3$ mol$^{-1}$ | 298.15 | $-1.902\ (\pm 0.016)$ | $-0.647\ (\pm 0.031)$ | $-0.369\ (\pm 0.072)$ | 0.007 |
| $\Delta\eta$ /mPa s | 298.15 | $-4.88\ (\pm 0.05)$ | $-17.3\ (\pm 0.3)$ | $4.75\ (\pm 0.19)$ | 0.010 |
| $\alpha_p^E$ /10$^{-6}$ K$^{-1}$ | 298.15 | $-59.8\ 10^{-6}$ | $-33.5\ 10^{-6}$ | $-1.3\ 10^{-6}$ | $1.2\ 10^{-7}$ |
|  |  | $(\pm 0.3\ 10^{-6})$ | $(\pm 0.6\ 10^{-6})$ | $(\pm 1.3\ 10^{-6})$ |  |
| $\Delta\left(\Delta G_m^*\right)$ /J mol$^{-1}$ | 298.15 | $-2634\ (\pm 22)$ | $-1254\ (\pm 43)$ | $-652\ (\pm 100)$ | 9 |
| $V_m^E$ /cm$^3$ mol$^{-1}$ | 303.15 | $-0.197\ (\pm 0.016)$ | $-0.663\ (\pm 0.032)$ | $-0.358\ (\pm 0.074)$ | 0.007 |
| $\Delta\eta$ /mPa s | 303.15 | $-4.055\ (\pm 0.013)$ | $-13.86\ (\pm 0.08)$ | $4.54\ (\pm 0.06)$ | 0.003 |
| $V_m^E$ /cm$^3$ mol$^{-1}$ | 308.15 | $-2.038\ (\pm 0.017)$ | $-0.685\ (\pm 0.034)$ | $-0.355\ (\pm 0.078)$ | 0.007 |
| $\Delta\eta$ /mPa s | 308.15 | $-3.410\ (\pm 0.019)$ | $-10.76\ (\pm 0.10)$ | $4.14\ (\pm 0.09)$ | 0.004 |

[a] $R = V_m^E$, $\alpha_p^E$, $\Delta\left(\Delta G_m^*\right)$, $\Delta\eta$ ; [b]Equation (11), units, the same as the corresponding property.

**Table 5.** Viscosity results provided by the application of the equations: Grunberg-Nissan (Eq. 3; adjustable parameter: $G_{12}$, Fang-He (Eq. 4, adjustable parameter, $W_{12}/RT$); and McAllister (Eq. 2; adjustable parameters: $Z_{12}$ and $Z_{21}$) and by the application of Bloomfield-Dewan's model to 1-iodonaphthalene (1) + n-alkane (2) mixtures at 298.15 K and atmospheric pressure.

| Grunberg-Nissan |  | Fang-He |  | McAllister |  |  | Bloomfield-Dewan |
|---|---|---|---|---|---|---|---|
| $G_{12}$ | $\sigma_r(\eta)$ [a] | $W_{12}/RT$ | $\sigma_r(\eta)$ [a] | $Z_{12}$ | $Z_{21}$ | $\sigma_r(\nu)$ [a] | $\sigma_r(\eta)$ [a] ( $\alpha=0$; $\beta=1$) |
|  |  | 1-iodonaphthalene (1) + heptane (2) |  |  |  |  |  |
| $-1.28$ | 0.03 | $-0.039$ | 0.025 | 1.252 | 0.799 | 0.009 | 0.049 |
|  |  | 1-iodonaphthalene (1) + decane (2) |  |  |  |  |  |
| $-1.54$ | 0.04 | $-0.029$ | 0.026 | 1.50 | 1.36 | 0.011 | 0.060 |
|  |  | 1-iodonaphthalene (1) + dodecane (2) |  |  |  |  |  |
| $-1.50$ | 0.04 | $-0.023$ | 0.027 | 1.80 | 1.92 | 0.011 | 0.043 |
|  |  | 1-iodonaphthalene (1) + tetradecane (2) |  |  |  |  |  |
| $-1.36$ | 0.04 | $-0.019$ | 0.024 | 2.23 | 2.68 | 0.009 | 0.024 |

[a]Equation (12).

## 5. Discussion

Below, values of the thermophysical properties of mixtures are provided at equimolar composition and 298.15 K. On the other hand, $n$ stands for the number of C atoms in the n-alkane.

### 5.1 Experimental data

1-Chloronaphthalene or 1-iodonaphthalene are similar chemicals with close values of their dipole moments (1.52 and 1.44 D, respectively [36]). In addition, they have also similar effective dipole moments ( $\bar{\mu}$ ). This quantity is useful to examine the impact of polarity on bulk properties, and it is defined by the equation [37,38]:

$$\bar{\mu} = \left[ \frac{\mu^2 N_A}{4\pi\varepsilon_0 V_m k_B T} \right]^{1/2} \qquad (13)$$

where all the symbols have the usual meaning. The rather low values of $\bar{\mu}$ (0.575 (1-chloronaphthalene); 0.455 (1-iodonaphthalene)) reveal that, in the mixtures formed by one of the mentioned compounds and $n$-alkanes, orientational effects are weak [5]. In fact, $H_m^E$ values of 1-chloronaphthalene + $n$-alkane mixtures are positive and not large (656 J mol$^{-1}$, for the system with heptane [19]). In view of these previous considerations, one can also expect positive values of $H_m^E$ for the corresponding systems with 1-iodonaphthalene. This is relevant since these solutions show negative values of $V_m^E$/cm$^3$ mol$^{-1}$: $-1.521$ ($n = 7$); $-0.817$ ($n = 10$); $-0.626$ ($n = 12$) and $-0.475$ ($n = 14$), and it means that the contribution from structural effects (that is, effects related to different shape, size or free volume of the mixture compounds) to $V_m^E$ is dominant. The negative values of $\frac{\Delta V_m^E}{\Delta T}$/cm$^3$ mol$^{-1}$ K$^{-1}$ ($-0.015$ ($n = 7$); $-0.006$ ($n = 10$); $-0.005$ ($n = 12$) and $-0.003$ ($n = 14$)) support this statement. The rather large and negative $\frac{\Delta V_m^E}{\Delta T}$ value obtained for the system with heptane is remarkable and indicates the existence of large structural effects in the solution. The mentioned effects can be of free volume type, as is shown by the large difference between the $\alpha_p$ /10$^{-3}$ K$^{-1}$ values: 0.652 (1-iodonaphthalene); 1.249 (heptane) (Table 2). Note that both compounds are of similar size (146.53 cm$^3$ mol$^{-1}$ (1-iodonaphthalene); 147.45 cm$^3$mol$^{-1}$ (heptane)). The observed increase of $V_m^E$ with $n$ may be ascribed to the existence of decreasing structural effects.

It is well known that positive values of $\Delta\eta$ are found in mixtures where strong interactions exist between unlike molecules, [28, 39-41]. Here, the values of ($\Delta\eta$ /mPa s) are negative ($-2.796$ ($n = 7$); $-2.498$ ($n = 10$); $-2.246$ ($n = 12$); $-1.972$ ($n = 14$) (Table 4)), which is a typical feature of systems where no specific interactions exist between the components [42,43]. It is quite clear that, in terms of viscosity, interactions between 1-iodonaphthalene molecules are largely disrupted by $n$-alkanes. Accordingly, the $\Delta\eta$ ($x_1$) curves are skewed towards large values of $x_1$ (Figure 3). In addition, $\Delta\eta$ also increases with $n$, i.e., the increase of the length of the alkane leads to a loss of the fluidization of the system.

It has been previously stated that a certain correlation exists between $\Delta\eta$ and $V_m^E$, and that these thermophysical properties have opposite signs [44,45]. This trend is not held for the present solutions since they are characterized by negative values of $\Delta\eta$ and $V_m^E$. We have already mentioned that the system with heptane shows large free volume effects. It is known that such effects lead to increased $\eta$ values [33]. However, our results of $\Delta\eta$ are very negative (see above, Table 3, Figure 3) and this suggests that free volume effects do not contribute meaningfully to $\Delta\eta$, a quantity currently determined, in large extent, by interactional effects.

Negative values of $\frac{\Delta(\Delta\eta)}{\Delta T}$ are encountered for systems where strong interaction between unlike molecules are dominant ($-0.021$ mPa s K$^{-1}$ for the methanol + cyclohexylamine mixture [28]), indicating that such interactions are broken when $T$ increases. In contrast, for the solutions with 1-iodonaphthalene, $\frac{\Delta(\Delta\eta)}{\Delta T}$ / mPa s K$^{-1}$ = 0.12 ($n = 7$); 0.11 ($n = 10$); 0.10 ($n = 12$); 0.09 ($n = 14$), which suggests that interactions between like molecules are dominant. That is, viscosity values approach those given by ($x_1\eta_1 + x_2\eta_2$) when $T$ increases, and this means that there is a lower change of the mixture fluidization. The same trend is observed for the ethanol + heptane [46], or methyl ester + $n$-alkane [47] mixtures.

### 5.2 Results from models

#### 5.2.1 Eyring's model

Firstly, we underline that the current results on $\Delta(\Delta G_m^*)$ are well correlated using Redlich-Kister expansions (Table 4). One can conclude then that our measurements at different temperatures are well performed. On the other hand, values of $\Delta G_m^*(x_1)$ show that the change of a molecule from an equilibrium position to a new one overcoming the attractive forces exerted by its neighbors is a process that depends mainly on enthalpic effects, since the values of $T\Delta S_m^*$ are ranged between $0.3\Delta G_m^*$ and $0.2\Delta G_m^*$ (Table S4). This effect becomes more important when $n$ is increased. In addition, at $x_1 = 0.5$, $\Delta G_m^*$ changes linearly with $n$ according to $\Delta G_m^*$/kJ mol$^{-1}$ = 12.5 + 0.406 $n$ ($r = 0.99996$), which remarks the importance of size effects on the relative variation of $\Delta G_m^*$ ($n$).

#### 5.2.2 Correlation equations

The application of the McAllister equation (two adjustable parameters) provides a mean relative standard deviation $\bar{\sigma}_r(\nu) = (1/N_S)\sum\sigma_r(\nu) = 0.010$ ($N_S$ is the number of systems, Tables 5 and S5), that is, there is an excellent agreement between experimental and calculated kinematic viscosities. Results are somewhat poorer when the Grunberg-Nissan equation is used (one adjustable parameter). For example, at 298.15 K, $\bar{\sigma}_r(\eta)$ (defined similarly to $\bar{\sigma}_r(\nu)$) is 0.037. The application of the equation (4) improves slightly this result ($\bar{\sigma}_r(\eta) = 0.025$). It seems that size effects are not very relevant on $\eta$ results for this type of system. It is important to note that the adjustable parameter of the Grunberg-Nissan equation is negative at any temperature (Tables 5 and S5). Typically, negative values of $G_{12}$ have been ascribed to interactions arising from dispersion London forces (dispersive interactions) are dominant [39,43]. We have demonstrated that systems containing one *n*-alkane and $C_6H_5X$ (X = Cl, Br, I) or 1,2,4-trichlorobenzene or 1-chloronaphthalene are characterized by orientational effects rather weak [5,6], and one can expect the same trend in systems with 1-iodonaphthalene. Negative values of $G_{12}$ have been obtained for systems with iodobenzene, or 1-chloronaphthalene, or 1,2,4-trichlorobenzene [6].

### 5.2.3 Bloomfield-Dewan's model

The free volume theory ($\alpha = 1; \beta = 0$) provides rather poor results for the systems under consideration. Thus, at 298.15 K, $\sigma_r(\eta) = 0.230$ ($n = 7$); 0.147 ($n = 10$); 0.123 ($n = 12$); 0.103 ($n = 14$). The larger deviation is obtained for the solution with heptane, characterized by strong free volume effects. In this case, theoretical viscosities are lower than the experimental results and it reveals that the model is not suitable for this type of system. Calculations performed assuming that only the residual part contributes to viscosity ($\alpha = 0; \beta = 1$) lead to results which are meaningfully improved: $\sigma_r(\eta) = 0.049$ ($n = 7$); 0.060 ($n = 10$); 0.043 ($n = 12$); 0.024 ($n = 14$). It is remarkable that this is attained using DISQUAC for the calculation of $\Delta G_m^{RES}$ with interaction parameters determined from low-pressure phase equilibria data.

### 5.4 Comparison with related mixtures

The replacement of iodobenzene by 1-iodonaphthalene in systems with a given *n*-alkane leads to more negative values of $V_m^E$ and $\Delta \eta$ (Figure 6). It must be noted that $V_m^E$ results for solutions with $C_6H_5I$ are positive from $n \geq 10$ [7] (Figure 6), while the corresponding $V_m^E$ values for systems involving 1-iodonaphthalene are negative. That is, structural effects are more relevant in the latter solutions (compare $\alpha_p /10^{-3} K^{-1} = 0.652$ (1-iodonaphthalene); 0.837 (iodobenzene) [7]). Consequently, $\dfrac{\Delta V_m^E}{\Delta T}$ ($n = 7$)/cm$^3$ mol$^{-1}$ K$^{-1} = -0.015$ (1-iodonaphthalene); $-0.006$ ($C_6H_5I$) [7]. Results on $\Delta \eta$ are much more negative for the mixtures with 1-iodonaphthalene. For example, $\Delta \eta$ ($n = 7$)/mPa s $= -0.284$ ($C_6H_5I$) [6]; $-2.796$ (1-iodonaphthalene). It is clear that the alkane breaks more easily interactions between 1-iodonaphthalene molecules, leading to a higher fluidization of the system. This effect increases with the temperature and $\dfrac{\Delta \Delta \eta}{\Delta T}$ (heptane)/mPa s K$^{-1}$ = 0.12 (1-iodonaphthalene); 0.005($C_6H_5I$) [6]. There is a lack of viscosity data for mixtures with $C_6H_5Cl$. Nevertheless, it seems that solutions with $C_6H_5Cl$ or 1-chloronaphthalene show similar trends to those stated above. Thus, $V_m^E$ ($n = 6$)/cm$^3$ mol$^{-1} = -0.521$ ($C_6H_5Cl$ [48]); $-1.560$ (1-chloronaphthalene) [19] and $\Delta \eta$ ($n = 6$)/mPa s $= -0.068$ ($C_6H_5Cl$, $T = 303.15$ K [49]); $-0.871$ (1-chloronaphthalene) [20]. That is, structural effects are stronger in solutions with 1-chloronaphthalene ($\alpha_p /10^{-3} K^{-1} = 0.99$ ($C_6H_5Cl$), 0.70 (1-chloronaphthalene) [5]), characterized also by higher fluidization. When comparing systems with 1-iodo or 1-chloronaphthalene, larger differences between $V_m^E$ results are encountered for solutions with heptane or decane (Figure 6) where structural effects are more relevant when 1-iodonaphthalene is involved. These solutions also show more negative $\Delta\eta$ results.

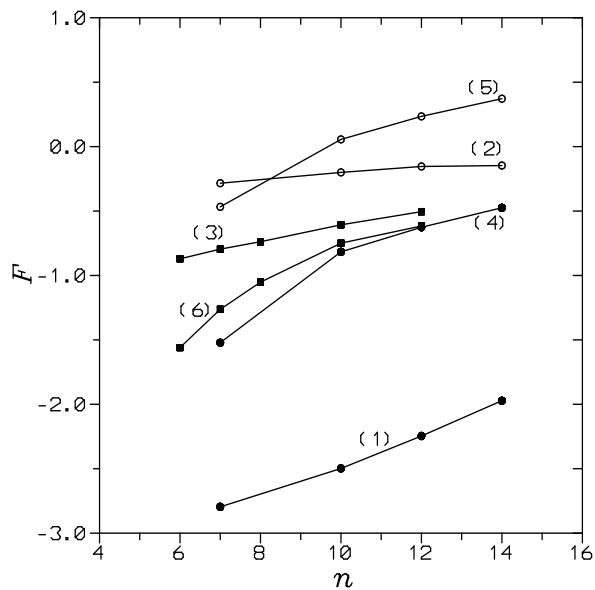

**Figure 6.** Thermophysical properties ($F = (V_m^E /cm^3 \ mol^{-1})$; $(\Delta\eta /mPa \ s))$ for $n$-alkane mixtures at equimolar composition, 298.15 K and atmospheric pressure. Points, experimental results. $\Delta\eta$ : (1), 1-iodonaphthalene + $n$-alkane (this work); (2), iodobenzene + $n$-alkane [7]; (3), 1-chloronaphthalene + $n$-alkane [20]. $V_m^E$ : (4), 1-iodonaphthalene + $n$-alkane (this work); (5), iodobenzene + $n$-alkane [7]; (6), 1-chloronaphthalene + $n$-alkane [19]. Lines are for the aid of the eye; $n$ is the number of C atoms in the $n$-alkane.

It is interesting to compare results from the Eyring's model for the solutions cited above. In the case of heptane systems, we have $\Delta H^* /kJ \ mol^{-1} = 8.3$ ($C_6H_5I$); 11.4 (1-iodonaphthalene); 8.9 (1-chloronaphthalene); $T\Delta S^* /kJ \ mol^{-1} = -5.1$ ($C_6H_5I$); $-3.9$ (1-iodonaphthalene), $-5.3$ (1-choronaphthalene). Thus, enthalpic effects become more relevant along the process in which a molecule of 1-iodonaphthalene flows from an equilibrium position to a new one.

## 6. Conclusions

Data on $\rho$, $\nu$ and $\eta$ for the mixtures 1-iodonaphthalene + $n$-alkane at (288.15-308.15) K and atmospheric pressure are provided. $V_m^E$ and $\Delta\eta$ results are large and negative. The former indicates the existence of structural effects in the studied systems, which is also supported by negative values of $\frac{\Delta V_m^E}{\Delta T}$. Results on $\Delta\eta$ reveal that the interactions between 1-iodonaphthalene molecules are broken by $n$-alkanes in large extent. Interactions are of dispersive type, and size or free volume effects have little impact on viscosity results, well represented by the absolute rate model using $\Delta G_{\mathrm{m}}^{\mathrm{RES}}$ values from DISQUAC.

### Acknowledgements

J. V. A.-L. would like to thank the Instituto de Corresponsabilidade pela Educação (ICE) – Brazil for his PhD scholarship. F. P. acknowledges the FPI grant PREP2022-000047 from MCIN/AEI/10.13039/501100011033/ and FEDER, UE.

### Funding

This work was carried out under Project PID2022-137104NA-I00, funded by MICIN/AEI/10.13039/501100011033/ and by FEDER, UE.

# SUPPLEMENTARY MATERIAL

# VOLUMETRIC AND VISCOSITY DATA OF 1-IODONAPHTHALENE + *n*-ALKANE MIXTURES AT (288.15-308.15) K

Luis Felipe Sanz, Juan Antonio González[*], Fernando Hevia, Daniel Lozano-Martín, João Victor Alves-Laurentino, Fatemeh Pazoki, Isaías García de la Fuente, José Carlos Cobos

G.E.T.E.F., Departamento de Física Aplicada, Facultad de Ciencias, Universidad de Valladolid, Paseo de Belén, 7, 47011 Valladolid, Spain

*Corresponding author, e-mail: jagl@termo.uva.es

**Table S1.** Kinematic ($\nu$) viscosities of pure compounds at temperature $T$ and 94 kPa. Comparison of experimental (Exp.) results with literature (Lit.) values.[a]

| $T$/K | 1-iodonaphthalene | | heptane | | decane | | dodecane | | tetradecane | |
|---|---|---|---|---|---|---|---|---|---|---|
| | | | | | $\nu$ /cSt | | | | | |
| | Exp. | Lit. | Exp. | Lit. | Exp. | Lit. | Exp. | Lit. | Exp. | Lit. |
| 288.15 | 6.370 | | 0.635 | 0.629[S1] | 1.360 | 1.364[S2] 1.350[S3] | 2.192 | 2.170 | 3.402 | 3.380[S2] |
| 293.15 | 5.332 | | 0.604 | 0.600[S4] 0.604[S7] | 1.263 | 1.254[S5] 1.258[S4] 1.260[S6] | 2.001 | 1.982[S5] 1.974[S4] 1.960[S6] 1.990[S10] | 3.054 | 3.020[S6] 3.069[S8] 3.052[S9] |
| 298.15 | 4.535 | | 0.576 | 0.574[S1] 0.572[S4] 0.577[S13] 0.582[S14,S15] | 1.178 | 1.169[S2] 1.172[S3] 1.168[S5] 1.186[S16] 1.172[S4] 1.183[S7] | 1.835 | 1.809[S2] 1.818[S5] 1.825[S4] | 2.759 | 2.753[S8,S11] 2.739[S12] |
| 303.15 | 3.910 | | 0.549 | 0.549[S1] 0.551[S13] | 1.102 | 1.096[S5] 1.100[S6] | 1.691 | 1.682[S5] 1.670[S6] 1.680[S0] | 2.507 | 2.480[S6] 2.512[S9] |
| 308.15 | 3.409 | | 0.525 | 0.522[S4] 0.528[S13] 0.521[S18] | 1.034 | 1.034[S2] | 1.565 | 1.557[S2] 1.558[S4] | 2.289 | 2.271[S9] 2.277[S17] |

[a]The uncertainties are: $u(T)$ = 0.02 K; $u(p)$ = 10 kPa. For kinematic viscosity, the relative combined expanded uncertainty (0.95 level of confidence) is $U_{rc}(\nu)$ = 0.020 .

**Table S2.** Densities ( $\rho$ ) and kinematic viscosities ( $\nu$ ) of 1-iodonaphthalene (1) + $n$-alkane (2) mixtures at temperature $T$ and 94 kPa vs. $x_1$, the mole fraction of 1-iodonaphthalene[a].

| $x_1$ | $\rho$ /g cm$^{-3}$ | $\nu$ /cSt | $\rho$ /g cm$^{-3}$ | $\nu$ /cSt | $\rho$ /g cm$^{-3}$ | $\nu$ /cSt |
|---|---|---|---|---|---|---|
| | | | 1-iodonaphthalene (1) + heptane (2) | | | |
| | $T$ /K = 288.15 | | $T$ /K = 293.15 | | $T$ /K = 298.15 | |
| 0 | 0.68797 | 0.635 | 0.68378 | 0.604 | 0.67954 | 0.576 |
| 0.0956 | 0.79228 | 0.679 | 0.78777 | 0.644 | 0.78324 | 0.611 |
| 0.1976 | 0.90337 | 0.760 | 0.89861 | 0.716 | 0.89386 | 0.677 |
| 0.2993 | 1.01338 | 0.882 | 1.00839 | 0.825 | 1.00345 | 0.775 |
| 0.3980 | 1.11932 | 1.051 | 1.11417 | 0.975 | 1.10909 | 0.911 |
| 0.4989 | 1.22745 | 1.288 | 1.22214 | 1.185 | 1.21693 | 1.096 |
| 0.5993 | 1.33392 | 1.622 | 1.32847 | 1.475 | 1.32315 | 1.353 |
| 0.6970 | 1.43588 | 2.099 | 1.43037 | 1.887 | 1.42486 | 1.703 |
| 0.7982 | 1.54104 | 2.861 | 1.53545 | 2.524 | 1.52987 | 2.248 |
| 0.8924 | 1.63761 | 4.021 | 1.63198 | 3.477 | 1.62635 | 3.051 |
| 1 | 1.74524 | 6.370 | 1.73959 | 5.332 | 1.73394 | 4.535 |
| | $T$ /K = 303.15 | | $T$ /K = 308.15 | | | |
| 0 | 0.67529 | 0.549 | 0.67100 | 0.525 | | |
| 0.0956 | 0.77870 | 0.582 | 0.77414 | 0.554 | | |
| 0.1976 | 0.88914 | 0.641 | 0.88425 | 0.608 | | |
| 0.2993 | 0.99856 | 0.730 | 0.99345 | 0.689 | | |
| 0.3980 | 1.10408 | 0.852 | 1.09879 | 0.798 | | |
| 0.4989 | 1.21180 | 1.015 | 1.20635 | 0.946 | | |
| 0.5993 | 1.31793 | 1.247 | 1.31233 | 1.149 | | |
| 0.6970 | 1.41934 | 1.548 | 1.41383 | 1.416 | | |
| 0.7982 | 1.52429 | 2.016 | 1.51872 | 1.822 | | |
| 0.8924 | 1.62072 | 2.697 | 1.61511 | 2.408 | | |
| 1 | 1.72828 | 3.910 | 1.72265 | 3.409 | | |
| | | | 1-iodonaphthalene (1) + decane (2) | | | |
| | $T$ /K = 288.15 | | $T$ /K = 293.15 | | $T$ /K = 298.15 | |
| 0 | 0.73367 | 1.360 | 0.72990 | 1.263 | 0.72612 | 1.178 |
| 0.0981 | 0.81144 | 1.393 | 0.80743 | 1.290 | 0.80340 | 1.200 |
| 0.1958 | 0.89233 | 1.466 | 0.88808 | 1.344 | 0.88383 | 1.246 |
| 0.2960 | 0.97987 | 1.560 | 0.97541 | 1.433 | 0.97093 | 1.322 |
| 0.3971 | 1.07297 | 1.715 | 1.06829 | 1.566 | 1.06360 | 1.438 |
| 0.4981 | 1.17097 | 1.935 | 1.16609 | 1.755 | 1.16120 | 1.600 |
| 0.5962 | 1.27154 | 2.236 | 1.26648 | 2.008 | 1.26140 | 1.817 |
| 0.6957 | 1.37907 | 2.659 | 1.37383 | 2.360 | 1.36859 | 2.119 |
| 0.7968 | 1.49443 | 3.311 | 1.48904 | 2.908 | 1.48363 | 2.572 |
| 0.8958 | 1.61380 | 4.341 | 1.60826 | 3.734 | 1.60272 | 3.254 |
| 1 | 1.74524 | 6.370 | 1.73959 | 5.332 | 1.73394 | 4.535 |
| | $T$ /K = 303.15 | | $T$ /K = 308.15 | | | |
| 0 | 0.72234 | 1.102 | 0.71854 | 1.034 | | |
| 0.0981 | 0.79938 | 1.120 | 0.79534 | 1.049 | | |
| 0.1958 | 0.87957 | 1.160 | 0.87531 | 1.083 | | |
| 0.2960 | 0.96645 | 1.226 | 0.96197 | 1.141 | | |
| 0.3971 | 1.05891 | 1.327 | 1.05421 | 1.230 | | |
| 0.4981 | 1.15631 | 1.466 | 1.15142 | 1.352 | | |
| 0.5962 | 1.25633 | 1.653 | 1.25125 | 1.513 | | |
| 0.6957 | 1.36335 | 1.918 | 1.35810 | 1.739 | | |
| 0.7968 | 1.47824 | 2.296 | 1.47285 | 2.074 | | |
| 0.8958 | 1.59719 | 2.869 | 1.59166 | 2.548 | | |
| 1 | 1.72828 | 3.910 | 1.72265 | 3.409 | | |

| | $T/K = 288.15$ | | $T/K = 293.15$ | | $T/K = 298.15$ | |
|---|---|---|---|---|---|---|
| 0 | 0.75255 | 2.192 | 0.74893 | 2.001 | 0.74529 | 1.835 |
| 0.0996 | 0.81912 | 2.186 | 0.81528 | 1.990 | 0.81143 | 1.824 |
| 0.1981 | 0.89006 | 2.212 | 0.88601 | 2.011 | 0.88195 | 1.837 |
| 0.2994 | 0.96877 | 2.278 | 0.96450 | 2.063 | 0.96023 | 1.880 |
| 0.3981 | 1.05178 | 2.395 | 1.04730 | 2.157 | 1.04281 | 1.960 |
| 0.4971 | 1.14199 | 2.563 | 1.13730 | 2.300 | 1.13261 | 2.077 |
| 0.5974 | 1.24158 | 2.826 | 1.23669 | 2.512 | 1.23179 | 2.255 |
| 0.6975 | 1.34971 | 3.189 | 1.34462 | 2.814 | 1.33952 | 2.512 |
| 0.7987 | 1.46968 | 3.733 | 1.46438 | 3.259 | 1.45909 | 2.874 |
| 0.8990 | 1.60036 | 4.629 | 1.59487 | 3.977 | 1.58939 | 3.457 |
| 1 | 1.74524 | 6.370 | 1.73959 | 5.332 | 1.73394 | 4.535 |
| | $T/K = 303.15$ | | $T/K = 308.15$ | | | |
| 0 | 0.74167 | 1.691 | 0.73803 | 1.565 | | |
| 0.0996 | 0.80759 | 1.680 | 0.80374 | 1.551 | | |
| 0.1981 | 0.87790 | 1.690 | 0.87383 | 1.559 | | |
| 0.2994 | 0.95595 | 1.722 | 0.95168 | 1.586 | | |
| 0.3981 | 1.03834 | 1.787 | 1.03386 | 1.640 | | |
| 0.4971 | 1.12793 | 1.890 | 1.12324 | 1.724 | | |
| 0.5974 | 1.22689 | 2.037 | 1.22201 | 1.852 | | |
| 0.6975 | 1.33442 | 2.254 | 1.32933 | 2.035 | | |
| 0.7987 | 1.45379 | 2.554 | 1.44849 | 2.287 | | |
| 0.8990 | 1.58390 | 3.030 | 1.57842 | 2.691 | | |
| 1 | 1.72828 | 3.910 | 1.72265 | 3.409 | | |

| | $T/K = 288.15$ | | $T/K = 293.15$ | | $T/K = 298.15$ | |
|---|---|---|---|---|---|---|
| 0 | 0.76631 | 3.402 | 0.76277 | 3.054 | 0.75923 | 2.759 |
| 0.0988 | 0.82361 | 3.340 | 0.81988 | 2.994 | 0.81615 | 2.697 |
| 0.1996 | 0.88763 | 3.284 | 0.88371 | 2.939 | 0.87979 | 2.646 |
| 0.2992 | 0.95726 | 3.268 | 0.95314 | 2.930 | 0.94902 | 2.634 |
| 0.3962 | 1.03208 | 3.307 | 1.02776 | 2.933 | 1.02344 | 2.653 |
| 0.4971 | 1.11842 | 3.396 | 1.11389 | 3.007 | 1.10936 | 2.701 |
| 0.5968 | 1.21390 | 3.552 | 1.20915 | 3.140 | 1.20440 | 2.791 |
| 0.6973 | 1.32192 | 3.831 | 1.31695 | 3.357 | 1.31197 | 2.967 |
| 0.7984 | 1.44490 | 4.256 | 1.43969 | 3.716 | 1.43449 | 3.255 |
| 0.8983 | 1.58366 | 4.959 | 1.57822 | 4.238 | 1.57279 | 3.683 |
| 1 | 1.74524 | 6.370 | 1.73959 | 5.332 | 1.73394 | 4.535 |
| | $T/K = 303.15$ | | $T/K = 308.15$ | | | |
| 0 | 0.75570 | 2.507 | 0.75217 | 2.289 | | |
| 0.0988 | 0.81244 | 2.445 | 0.80872 | 2.232 | | |
| 0.1996 | 0.87588 | 2.400 | 0.87197 | 2.183 | | |
| 0.2992 | 0.94491 | 2.381 | 0.94080 | 2.173 | | |
| 0.3962 | 1.01913 | 2.386 | 1.01482 | 2.161 | | |
| 0.4971 | 1.10484 | 2.422 | 1.10031 | 2.196 | | |
| 0.5968 | 1.19966 | 2.498 | 1.19491 | 2.257 | | |
| 0.6973 | 1.30700 | 2.637 | 1.30204 | 2.368 | | |
| 0.7984 | 1.42929 | 2.861 | 1.42409 | 2.556 | | |
| 0.8983 | 1.56736 | 3.224 | 1.56194 | 2.862 | | |
| 1 | 1.72828 | 3.910 | 1.72265 | 3.409 | | |

**Table S3.** Excess isobaric thermal expansion coefficients ($\alpha_p^{\mathrm{E}}$) for 1-iodonaphthalene (1) + $n$-alkane (2) mixtures at 298.15 $K$ and 94 kPa, vs. $x_1$, the mole fraction of 1-iodonaphthalene[a].

| $x_1$ | $\alpha_p^{\mathrm{E}} / 10^{-6}\,\mathrm{K}^{-1}$ | $x_1$ | $\alpha_p^{\mathrm{E}} / 10^{-6}\,\mathrm{K}^{-1}$ | $x_1$ | $\alpha_p^{\mathrm{E}} / 10^{-6}\,\mathrm{K}^{-1}$ | $x_1$ | $\alpha_p^{\mathrm{E}} / 10^{-6}\,\mathrm{K}^{-1}$ |
|---|---|---|---|---|---|---|---|
| heptane | | decane | | dodecane | | tetradecane | |
| 0.0956 | $-33.85$ | 0.0981 | $-10.64$ | 0.0996 | $-5.10$ | 0.0988 | $-3.01$ |
| 0.1952 | $-63.72$ | 0.1958 | $-19.16$ | 0.1981 | $-9.84$ | 0.1996 | $-6.42$ |
| 0.2962 | $-80.39$ | 0.2960 | $-26.82$ | 0.2994 | $-14.99$ | 0.2992 | $-9.70$ |
| 0.3937 | $-89.35$ | 0.3971 | $-31.59$ | 0.3981 | $-19.03$ | 0.3962 | $-12.74$ |
| 0.4949 | $-88.27$ | 0.4981 | $-33.86$ | 0.4971 | $-21.59$ | 0.4971 | $-14.89$ |
| 0.5947 | $-79.98$ | 0.5962 | $-33.10$ | 0.5974 | $-22.51$ | 0.5968 | $-15.73$ |
| 0.6970 | $-59.65$ | 0.6957 | $-30.03$ | 0.6975 | $-21.08$ | 0.6973 | $-15.67$ |
| 0.7982 | $-43.20$ | 0.7968 | $-23.72$ | 0.7987 | $-16.65$ | 0.7984 | $-12.90$ |
| 0.8924 | $-24.38$ | 0.8958 | $-13.84$ | 0.8990 | $-9.57$ | 0.8983 | $-7.89$ |

[a]The uncertainties are: $u(T) = 0.02$ K; $u(p) = 10$ kPa; $u(x_1) = 0.0010$. The relative combined expanded uncertainty (0.95 level of confidence) for $\alpha_p^{\mathrm{E}}$ is $U_{\mathrm{rc}}\left(\alpha_p^{\mathrm{E}}\right) = 0.04$.

**Table S4.** Parameters of activation of viscous flow: Gibbs energy $\Delta G^*$, enthalpy $\Delta H^*$, entropy $\Delta S^*$ and $\Delta\left(\Delta G^*\right)$ for 1-iodonaphthalene (1) + $n$-alkane (2) mixtures at 298.15 K and 94 kPa $vs.$ $x_1$, the mole fraction of 1-iodonaphthalene.

| $x_1$ | $\Delta H^*$ /kJ mol⁻¹ | $\Delta S^*$ /kJ K⁻¹ mol⁻¹ | $\Delta G^*$ /kJ mol⁻¹ | $\Delta\left(\Delta G^*\right)$ /kJ mol⁻¹ |
|---|---|---|---|---|
| | | 1-iodonaphthalene (1) + heptane (2) | | |
| 0 | 7.0 | − 17.8 | 12.3 | |
| 0.0956 | 7.5 | − 17.9 | 12.8 | − 223 |
| 0.1976 | 8.2 | − 17.4 | 13.4 | − 412 |
| 0.2993 | 9.1 | − 16.3 | 14.0 | − 552 |
| 0.3980 | 10.1 | − 15.2 | 14.6 | − 642 |
| 0.4989 | 11.4 | − 13.2 | 15.3 | − 707 |
| 0.5993 | 12.7 | − 11.4 | 16.1 | − 721 |
| 0.6970 | 14.5 | − 7.6 | 16.8 | − 689 |
| 0.7982 | 16.7 | -3.5 | 17.7 | − 572 |
| 0.8924 | 18.9 | 1 | 18.6 | − 357 |
| 1 | 23.1 | 11.0 | 19.8 | |
| | | 1-iodonaphthalene (1) + decane (2) | | |
| 0 | 10.1 | − 16.3 | 15.0 | |
| 0.0981 | 10.4 | − 15.9 | 15.2 | − 240 |
| 0.1958 | 10.9 | − 15.3 | 15.5 | − 445 |
| 0.2960 | 11.5 | − 14.2 | 15.8 | − 613 |
| 0.3971 | 12.3 | − 13.0 | 16.2 | − 735 |
| 0.4981 | 13.3 | − 11.1 | 16.6 | − 809 |
| 0.5962 | 14.4 | − 8.7 | 17.0 | − 831 |
| 0.6957 | 15.6 | − 6.4 | 17.5 | − 795 |
| 0.7968 | 17.3 | − 2.7 | 18.1 | − 673 |
| 0.8958 | 19.6 | 2.7 | 18.8 | − 446 |
| 1 | 23.3 | 11.7 | 19.9 | |
| | | 1-iodonaphthalene (1) + dodecane (2) | | |
| 0 | 12.4 | − 13.7 | 16.5 | |
| 0.0996 | 12.6 | − 13.4 | 16.6 | − 221 |
| 0.1981 | 12.9 | − 13.0 | 16.8 | − 409 |
| 0.2994 | 13.4 | − 11.9 | 16.9 | − 571 |
| 0.3981 | 14.0 | − 10.6 | 17.1 | − 688 |
| 0.4971 | 14.6 | − 9.2 | 17.4 | − 764 |
| 0.5974 | 15.6 | − 7.0 | 17.7 | − 789 |
| 0.6975 | 16.5 | − 5.0 | 18.0 | − 755 |
| 0.7987 | 18.1 | − 1.3 | 18.5 | − 657 |
| 0.8990 | 20.0 | 3.5 | 19.0 | − 436 |
| 1 | 23.1 | 11.0 | 19.8 | |
| | | 1-iodonaphthalene (1) + tetradecane (2) | | |
| 0 | 14.6 | − 11.0 | 17.9 | |
| 0.0988 | 14.9 | − 10.2 | 17.9 | − 171 |
| 0.1996 | 15.1 | − 9.7 | 17.9 | − 339 |
| 0.2992 | 15.1 | − 9.7 | 18.0 | − 472 |
| 0.3962 | 15.6 | − 8.2 | 18.1 | − 583 |
| 0.4971 | 16.1 | − 7.1 | 18.2 | − 658 |
| 0.5968 | 16.8 | − 5.2 | 18.3 | − 695 |
| 0.6973 | 17.8 | − 2.5 | 18.5 | − 672 |
| 0.7984 | 18.9 | 0.3 | 18.8 | − 576 |
| 0.8983 | 20.3 | 3.6 | 19.20 | − 389 |
| 1 | 23.1 | 11.0 | 19.8 | |

**Table S5.** Results provided by the application of the Grunberg-Nissan model (Eq. 3; adjustable parameter: ($G_{12}$) and McAllister (Eq. 2; adjustable parameters: $Z_{12}$ and $Z_{21}$) to 1-iodonaphthalene (1) + $n$-alkane (2) mixtures at temperature $T$ and 94 kPa.

| Equation | | $T$ = 288.15 K | | $T$ = 293.15 K | | $T$ = 298.15 K | | $T$ = 303.15 K | | $T$ = 308.15 K | |
|---|---|---|---|---|---|---|---|---|---|---|---|
| | | Param.[a] | $\sigma_r$[b] | Param.[a] | $\sigma_r$[b] | Param.[a] | $\sigma_r$[b] | Param.[a] | $\sigma_r$[b] | Param.[a] | $\sigma_r$[b] |
| | | 1-iodonaphthalene (1) + heptane (2) | | | | | | | | | |
| Grunberg-Nissan | $G_{12}$ | $-1.59$ | 0.05 | $-1.42$ | 0.04 | $-1.28$ | 0.03 | $-1.16$ | 0.03 | $-1.05$ | 0.02 |
| McAllister | $Z_{12}$ | 1.424 | 0.013 | 1.333 | 0.012 | 1.252 | 0.009 | 1.172 | 0.008 | 1.101 | 0.006 |
| | $Z_{21}$ | 0.932 | | 0.861 | | 0.799 | | 0.747 | | 0.698 | |
| | | 1-iodonaphthalene (1) + decane (2) | | | | | | | | | |
| Grunberg-Nissan | $G_{12}$ | $-1.83$ | 0.06 | $-1.67$ | 0.05 | $-1.54$ | 0.04 | $-1.42$ | 0.04 | $-1.32$ | 0.03 |
| McAllister | $Z_{12}$ | 1.74 | 0.015 | 1.61 | 0.013 | 1.50 | 0.011 | 1.40 | 0.009 | 1.300 | 0.008 |
| | $Z_{21}$ | 1.64 | | 1.49 | | 1.36 | | 1.24 | | 1.15 | |
| | | 1-iodonaphthalene (1) + dodecane (2) | | | | | | | | | |
| Grunberg-Nissan | $G_{12}$ | $-1.76$ | 0.05 | $-1.63$ | 0.05 | $-1.50$ | 0.04 | $-1.41$ | 0.04 | $-1.32$ | 0.04 |
| McAllister | $Z_{12}$ | 2.14 | 0.014 | 1.96 | 0.013 | 1.80 | 0.011 | 1.65 | 0.011 | 1.52 | 0.009 |
| | $Z_{21}$ | 2.37 | | 2.12 | | 1.92 | | 1.75 | | 1.60 | |
| | | 1-iodonaphthalene (1) + tetradecane (2) | | | | | | | | | |
| Grunberg-Nissan | $G_{12}$ | $-1.58$ | 0.05 | $-1.46$ | 0.04 | $-1.36$ | 0.04 | $-1.29$ | 0.04 | $-1.21$ | 0.03 |
| McAllister | $Z_{12}$ | 2.71 | 0.011 | 2.46 | 0.009 | 2.23 | 0.009 | 2.02 | 0.008 | 1.85 | 0.007 |
| | $Z_{21}$ | 3.37 | | 2.98 | | 2.68 | | 2.41 | | 2.18 | |

[a]Adjustable parameter; [b]standard relative deviation (Eq. 12).